\documentclass[preprint]{aastex63}
\usepackage{amssymb}
\usepackage{ulem}
\newcommand{\FeHle}[1]{$\mbox{[Fe/H]}\le{#1}$}   % [Fe/H] <= X
\newcommand{\tefft}{$T_{\mbox{\scriptsize eff}}$} % Command for Tefft, text mode
\newcommand{\teffm}{T_{\mbox{\scriptsize eff}}}  % Command for Tefft, math mode
\newcommand{\Tefft}{\ensuremath{T_{\mathrm{eff}}}}

\received{}
\revised{}
\accepted{March 2022}
\submitjournal{ApJ}

\shorttitle{Metal-Poor Stars studied with LAMOST/Subaru}
\shortauthors{Aoki et al.}
\begin{document}

\title{Four-hundred Very Metal-poor Stars studied with LAMOST and Subaru. I. Survey Design, 
  Follow-up Program, and Binary Frequency
\footnote{}}

\correspondingauthor{Wako Aoki, Gang Zhao}
\email{aoki.wako@nao.ac.jp, gzhao@nao.cas.cn}

\author[0000-0002-8975-6829]{Wako Aoki}
\affiliation{National Astronomical Observatory of Japan \\ 2-21-1 Osawa, Mitaka, Tokyo 181-8588, Japan}
\affiliation{Department of Astronomical Science, School of Physical Sciences, The Graduate University of Advanced Studies (SOKENDAI) \\
 2-21-1 Osawa, Mitaka, Tokyo 181-8588, Japan}

\author[0000-0002-0786-7307]{Haining Li}
\affiliation{Key Lab of Optical Astronomy, National Astronomical Observatories, Chinese Academy of Sciences (CAS) \\
A20 Datun Road, Chaoyang, Beijing 100101, China}
  
\author[0000-0002-8077-4617]{Tadafumi Matsuno}
\affiliation{Kapteyn Astronomical Institute, University of Groningen  \\ Landleven 12, 9747 AD Groningen, The Netherlands}
\affiliation{Department of Astronomical Science, School of Physical Sciences, The Graduate University of Advanced Studies (SOKENDAI) \\
 2-21-1 Osawa, Mitaka, Tokyo 181-8588, Japan}
\affiliation{National Astronomical Observatory of Japan \\ 2-21-1 Osawa, Mitaka, Tokyo 181-8588, Japan}

\author{Qianfan Xing}
\affiliation{Key Lab of Optical Astronomy, National Astronomical Observatories, Chinese Academy of Sciences (CAS) \\
A20 Datun Road, Chaoyang, Beijing 100101, China}

\author{Yuqin Chen}
\affiliation{Key Lab of Optical Astronomy, National Astronomical Observatories, Chinese Academy of Sciences (CAS) \\
A20 Datun Road, Chaoyang, Beijing 100101, China}

\author[0000-0002-4043-2727]{Norbert Christlieb}
\affiliation{Zentrum f\"ur Astronomie der Universit\"at Heidelberg, Landessternwarte, K\"onigstuhl 12, 69117 Heidelberg, Germany}

\author{Satoshi Honda}
\affiliation{Nishi-Harima Astronomical Observatory, Center for Astronomy, University of Hyogo \\ 407-2,
Nishigaichi, Sayo-cho, Sayo, Hyogo 679-5313, Japan}

\author[0000-0003-4656-0241]{Miho N. Ishigaki}
\affiliation{National Astronomical Observatory of Japan \\ 2-21-1 Osawa, Mitaka, Tokyo 181-8588, Japan}

\author[0000-0002-0349-7839]{Jianrong Shi}
\affiliation{Key Lab of Optical Astronomy, National Astronomical Observatories, Chinese Academy of Sciences (CAS) \\
A20 Datun Road, Chaoyang, Beijing 100101, China}

\author[0000-0002-4318-8715]{Takuma Suda}
\affiliation{Department of Liberal Arts, Tokyo University of Technology, Nishi Kamata 5-23-22, Ota-ku, Tokyo 144-8535, Japan
}

\author[0000-0001-8537-3153]{Nozomu Tominaga}
\affiliation{National Astronomical Observatory of Japan \\ 2-21-1 Osawa, Mitaka, Tokyo 181-8588, Japan}
\affiliation{Department of Astronomical Science, School of Physical Sciences, The Graduate University of Advanced Studies (SOKENDAI) \\
 2-21-1 Osawa, Mitaka, Tokyo 181-8588, Japan}
\affiliation{Department of Physics, Faculty of Science and Engineering, Konan University \\ 8-9-1 Okamoto, Kobe, Hyogo 658-8501, Japan
}
\affiliation{Kavli Institute for the Physics and Mathematics of the
Universe (WPI), The University of Tokyo, 5-1-5 Kashiwanoha, Kashiwa, Chiba
277-8583, Japan
}

\author[0000-0002-8609-3599]{Hong-Liang Yan}
\affiliation{Key Lab of Optical Astronomy, National Astronomical Observatories, Chinese Academy of Sciences (CAS) \\
A20 Datun Road, Chaoyang, Beijing 100101, China}
\affiliation{School of Astronomy and Space Science, University of Chinese Academy of Sciences \\
No.19(A) Yuquan Road, Shijingshan District, Beijing, 100049, China}

\author{Jingkun Zhao}
\affiliation{Key Lab of Optical Astronomy, National Astronomical Observatories, Chinese Academy of Sciences (CAS) \\
A20 Datun Road, Chaoyang, Beijing 100101, China}

\author[0000-0002-8980-945X]{Gang Zhao}
\affiliation{Key Lab of Optical Astronomy, National Astronomical Observatories, Chinese Academy of Sciences (CAS) \\
A20 Datun Road, Chaoyang, Beijing 100101, China}
\affiliation{School of Astronomy and Space Science, University of Chinese Academy of Sciences \\
No.19(A) Yuquan Road, Shijingshan District, Beijing, 100049, China}



\begin{abstract}

The chemical abundances of very metal-poor stars provide important 
constraints on the nucleosynthesis of the first generation of stars and early
chemical evolution of the Galaxy. We have obtained high-resolution
spectra with the Subaru Telescope for candidates of very metal-poor stars selected with a 
large survey of Galactic stars carried out with LAMOST. In this series of papers,
we report on the elemental abundances of about 400 very metal-poor
stars and discuss the kinematics of the sample obtained by combining 
the radial velocities measured in this study and recent astrometry
obtained with {\it Gaia}. This paper provides an overview of our
survey and follow-up program, and reports radial velocities for the whole sample.  We
identify seven double-lined spectroscopic binaries from our
high-resolution spectra, for which radial velocities of the
components are reported. We discuss the frequency of such relatively
short-period binaries at very low metallicity.

\end{abstract}



\section{Introduction} \label{sec:intro}

Very metal-poor stars found in solar-neighborhood are low-mass old
stars that provide a record of the chemical composition and dynamics of
the early Galaxy. For instance, nucleosynthesis yields of the first
generation of massive stars are believed to be preserved in the most
metal-poor stars, which enable us to estimate the masses of these
stars \citep[e.g., ][]{Heger2010ApJ, Nomoto2013ARAA}. The trend and
scatter of abundance ratios of the elements (e.g. $\alpha$/Fe ratios), as
well as the metallicity distribution, are strong constraints on the
formation history of the Galaxy \citep{Freeman2002ARAA}. Low-mass star
formation in the early Galaxy and their evolution are studied based on chemical abundances of a large sample of
very metal-poor stars, which were born as low-mass stars from
low-metallicity gas clouds \citep{Frebel2015ARAA}.

Searches for stars that record the products of early generations
of stars require a large survey of candidates of metal-poor stars, because such objects are quite
rare.  In the past two decades, continuous efforts of spectroscopic
studies have provided chemical abundance data based on high-resolution spectra for several hundreds of very
metal-poor stars \citep[e.g., ][]{Cayrel2004AA, Cohen2004ApJ, Honda2004ApJ, Lai2008ApJ, Cohen2013ApJ}. Among them, \citet{Barklem2005AA} determined metallicity
and elemental abundance ratios based on high-resolution, moderate  signal-to-noise ratio (SNR) spectra 
for 253 very metal-poor stars selected from the Hamburg/ESO survey \citep{Christlieb2008AA}. 
\citet{Barklem2005AA} demonstrate that such approach based on  high-resolution, short exposure spectra, 
called  ``snap-shot'' spectroscopy, is very efficient to investigate the overall abundance distributions of metal-poor stars 
in which absorption lines are very weak in general. 

Following the achievements in the early 2000s, recent
spectroscopic studies for very metal-poor stars have been showing rapid
progress by efficient follow-up observations for
targets found by new surveys. For instance, \citet{Aoki2013AJ} applied
this approach to very metal-poor stars selected from SDSS \citep{Yanny2009AJ}, providing a homogeneous set
of chemical abundance data for 137 objects, most of which are main-sequence turn-off stars. The data volume and their
homogeneity are useful to study the statistics not only of chemical
abundance ratios, but also of other stellar properties like binary
frequency, and also to select extreme objects for further
observations like the $\alpha$-poor object SDSS~J0018--0939 \citep{Aoki2014Sci}. 
Ultra metal-poor stars ([Fe/H]$<-4$) have been 
found in the SDSS sample by further studies, including \citet{Bonifacio2015AA}. Spectroscopic follow-up of metal-poor star candidates
found in the photometric Skymapper survey reported the most
metal-poor stars with and without detection of Fe 
\citep{Nordlander2019MNRAS, Keller2014Nature} as well as abundance distributions of many metal-poor stars \citep{Jacobson2015ApJ, Yong2021}. 
Very metal-poor stars have also been discovered by the Pristine photometric survey and follow-up spectroscopic studies \citep{Venn2020MNRAS}.

In addition to these new surveys, \citet{Yong2013ApJ} studied chemical abundances of 190 very metal-poor stars, 
most of which are obtained by reanalyses of spectra obtained by previous
studies, providing abundance results based on a homogeneous analysis. 

Very large surveys with high-resolution ($R\sim 20,000-30,000$) multi-object spectroscopy have been conducted in the last decade, 
providing huge data sets of chemical abundances for Galactic stars (e.g., Gaia/ESO survey: \cite{Gilmore2012Msngr}; GALAH survey: \cite{DeSilva2015MNRAS}; APOGEE: \cite{Majewski2017AJ}). The sample size is several tens or a hundred  thousands of field and cluster stars. 
The majority of these samples, however, are Galactic thin and thick disk stars reflecting the populations of stars in the solar neighbourhood. 
Moreover, elemental abundance ratios are not well determined for metal-poor stars due to the small number of detectable spectral lines for metal-poor stars 
in the limited wavelength ranges of the multi-object spectroscopy optimized for disk stars. 

A large sample of candidates of very metal-poor stars has been
obtained with LAMOST (Large sky Area Multi-Object fiber Spectroscopic Telescope)
\footnote{\url{http://www.lamost.org/public/}}. With an effective aperture of 4m, LAMOST is
capable of obtaining up to 4000 spectra simultaneously in a 5 degree diameter
field of view. The spectra cover the wavelength region 3700--9100~{\AA} with $R\sim 1800$ \citep{Zhao2012RAA}.  

The first regular, five-year survey, completed in 2017 (LAMOST-I), has obtained
about 9 million spectra of Galactic stars \citep{Yan2022Innovation}.

An advantage of the target selection in the LAMOST
  survey is that there is no selection bias on spectral type (or
  color). 
  This is quite
different from the survey of Galactic stars in SDSS (SEGUE), in which
targets for the medium resolution spectroscopy were pre-selected
\citep{Yanny2009AJ}.  Another advantage of the LAMOST
survey is that it covers relatively bright stars ($g \lesssim 15$) in a 
very wide area of the sky ($\sim 11,000$~deg$^{2}$) accessible from
the northern hemisphere. This provides a large sample of bright 
candidates for very metal-poor stars to which follow-up high
resolution spectroscopy can be applied with a reasonable amount of telescope
time. 

Follow-up observations for the selected candidates have been conducted with the Subaru Telescope High Dispersion Spectrograph \citep[HDS: ][]{Noguchi2002PASJ} to determine elemental abundances and obtain other stellar properties such as radial velocities, stellar parameters. The first attempt of follow-up observation in 2013 for candidates selected from the early LAMOST survey in 2012 was not very successful due to the uncertainty of metallicity estimates from the LAMOST spectra with limited SNR. With a significant improvement of the data quality of the LAMOST spectra and the pipeline to derive stellar parameters, the {\it Normal Program} of the Subaru Telescope conducted in 2014 successfully obtained useful high-resolution spectra for more than 50 metal-poor stars, confirming the high efficiency of candidate selection. Early results of our studies obtained from this first run were published by \citet{Li2015PASJ, Li2015RAA}. Following two other observing runs in {\it Normal Programs}, we were awarded an {\it Intensive Program} using 20 nights of the Subaru Telescope in 2016--2018. In the course of these observing programs, we have obtained "snap-shot" spectra for more than 400 very metal-poor stars with sufficient quality to determine chemical abundance ratios. A small portion of the observing time has been used to obtain high SNR spectra in the blue wavelength region to study detailed chemical abundances of extremely metal-poor stars found from the snap-shot spectra. Using these spectra, the detailed abundances of carbon-enhanced, extremely/ultra metal-poor stars were studied by \citet{Aoki2018PASJ} and \citet{Zhang2019PASJ}.

The observing time of the {\it Intensive Program} has been partially applied to small samples of metal-poor stars selected by different criteria. One is a sample of candidates of moving group stars having similar kinematics properties found based on the LAMOST survey, which could be remnants of dissolved clusters or dwarf galaxies. The result obtained for this sample was reported by \citet{Zhao2018ApJ} and \citet{Liang2018ApJ}. Another small sample is selected based on the $\alpha$/Fe ratios estimated from LAMOST spectra to study metal-poor stars with significantly low abundances of $\alpha$-elements, which might be related to the chemical evolution of small stellar systems like dwarf galaxies and/or specific nucleosynthesis of first-generation very massive objects. The first result obtained for a unique low-$\alpha$ star with very large r-process-excess was reported by \citet{Xing2019NatAs}. 
Through the observing runs, bright candidates of Li-enhanced red giants, regardless metallicity, have been observed using the time with insufficient weather condition for observations of fainter targets and twilight. The data of these observations are partially used in the study by \citet{Yan2021NatAs}.

This series of papers reports on the chemical abundances and other stellar properties including stellar parameters and radial velocities of the main sample of these programs. The sample consists of more than 400 very metal-poor stars. This is the first paper of the series, describing the sample selection (\S~2), high-resolution spectroscopy and data reduction (\S~3), radial velocity measurements (\S~4), and estimates of interstellar absorption from \ion{Na}{1} D lines. Through this study, seven objects are identified to be double-lined spectroscopic binaries having pairs of absorption lines in the high-resolution spectra. The binary property and metallicity estimated for these objects are reported in \S~6. 

Detailed chemical abundances and stellar parameters obtained from the high-resolution spectra for the main sample are reported in Li et al. (in preparation; Paper II). The sample of this paper and Paper II includes Li-enhanced very metal-poor stars reported by \citet{Li2018ApJS}. On the other hand, the metal-poor post-AGB star CC Lyr, identified as LAMOST~J~1833+3138 \citep{Aoki2017PASJ} is excluded from the sample.

\section{Sample Selection from LAMOST Survey}\label{sec:selection}

\subsection{Selection of candidates of very metal-poor stars}\label{subsec:LAMOST_selection}

Candidates of very metal-poor stars were selected from LAMOST DR1 through DR5 
for the follow-up program conducted with the
Subaru Telescope in 2014-2017. LAMOST provides low-resolution spectra
($R\sim 1800$) for the full optical wavelength range (3700--9100~{\AA}). 
We adopted a template matching method
to derive the metallicity of the program stars from LAMOST low-resolution spectra. Objects with sufficient data quality, 
e.g., SNR higher than 10 and 15 in the $g-$ and $r-$bands, respectively, are selected for this purpose. 

Two methods are used to determine the metallicity of each star, 
both based on comparisons with a grid of synthetic spectra 
adopting the ATLAS9 grid of stellar model atmospheres of \citet{Castelli&Kurucz2003IAUS}. 
The grid of synthetic spectra covers $4000\,\mathrm{K}\le\teffm\le 9000\,\mathrm{K}$,
$0.0 \le\log g \le 5.0$, and $-4.5\le\mathrm{[Fe/H]}\le-0.5$.
The first method is based on a direct comparison of normalized observed flux and synthetic spectra
in the wavelength range from 4360\,{\AA} to 5500\,{\AA} 
that includes sufficient numbers of absorption-line features
that are sensitive to stellar parameters, especially for low-metallicity stars, 
avoiding possible contamination from the CH $G-$band around 4300\,{\AA}.
The second method makes use of 27 line indices including, e.g., the \ion{Ca}{2}~K line index, 
which match the observed sets of line indices for the program stars
with the synthetic sets to find the best-fit stellar parameters.
When both methods have derived metallicities of \FeHle{-2.0}, 
and effective temperatures in the range of 4000\,K $<$ {\Tefft} $<$ 7000\,K, 
the object was then considered as a preliminary candidate.
Spectra of candidates selected with these criteria
are visually inspected to remove false positives such as cool white dwarfs, 
or stars that were selected because their spectra are disturbed by reduction artifacts or too low SNR. 
The whole procedure has resulted in over 15,000 candidate VMP stars, which has been used for further selection of follow-up observations.
For more details about the properties of LAMOST data, as well as the methodology of VMP star candidate selection from LAMOST spectra, 
readers may refer to \citet{Li2018ApJS}, 
which has adopted the same procedure on the low-resolution spectra from LAMOST DR3, 
and resulted in about 10,000 VMP candidates that are also included in the above mentioned preliminary candidate list.

\subsection{Selection of targets for Subaru follow-up}

Our programs of high-resolution spectroscopy for selected metal-poor
star candidates were conducted in 2014-2017 (see next section). 
We started the target selection for Subaru follow-up observations 
from a list of over 1500 very/extremely metal-poor star candidates 
which have been selected from LAMOST DR1 through DR5 as described in \S~\ref{subsec:LAMOST_selection}, 
and suitable for observations during our Subaru runs.
The primary targets of the programs in the first two years were extremely
metal-poor stars, and, hence, higher priority was given to candidates for which extremely
low-metallicity is indicated by the above estimates. 
The observing program in the remaining two years, i.e. the intensive program (\S~1), extends the study to
metallicity up to [Fe/H]$=-2.0$ for more comprehensive investigation
of early chemical enrichment in the Galaxy. 
For this purpose, we made a list of bright metal-poor star candidates %($g\lesssim 13.5$) 
including over 300 objects with $g\lesssim 13.5$,
and have randomly conducted high-resolution spectroscopy regardless of the
estimated metallicity. In addition to this sample, 
about 150 candidates of extremely metal-poor stars with $13.5<g<15$ 
were also selected as targets for the program. 

%From this procedure, about 1000 stars were selected as candidates for follow-up high-resolution spectroscopy. 

\begin{figure}
% \begin{center}
%  \includegraphics[width=10cm]{v0hist.eps}%{c1v.eps}
% \end{center}
\plotone{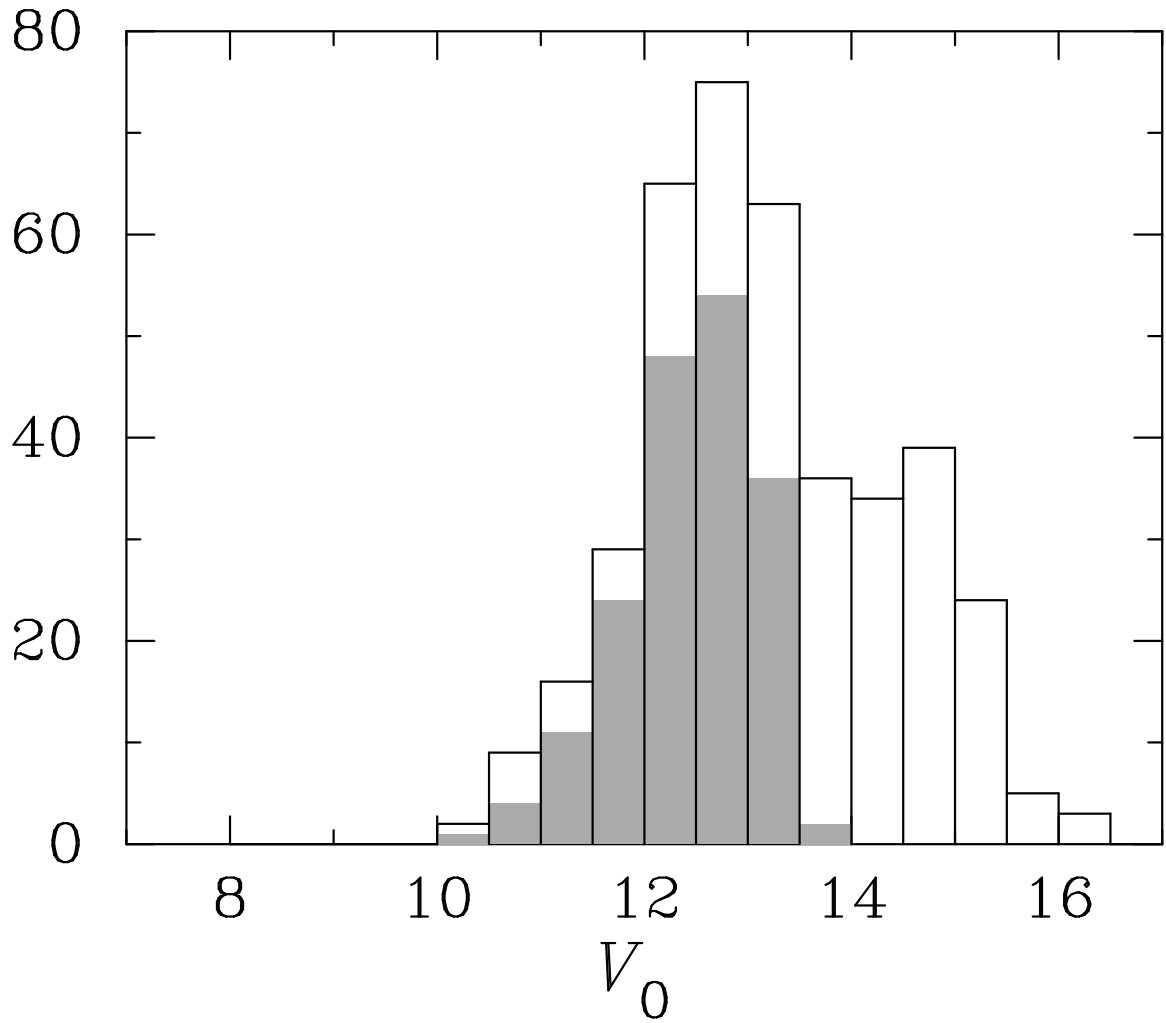}
 \caption{Distribution of the $V_{0}$ magnitude of our whole sample. The gray histogram presents the distribution of bright metal-poor star sample (see text)}.\label{fig:v0hist}
\end{figure}

In Table~\ref{tab:obj}, we provide photometry and reddening data for our sample. The photometry data are taken from APASS for optical bands, and 2MASS \citep{Skrutskie2006AJ} for J and K. The reddening data are adopted from \citet{Green2018MNRAS}. 
The distribution of the $V_{0}$ magnitudes is shown in
Figure~\ref{fig:v0hist}. The gray histogram presents the $V_{0}$ distribution of the bright metal-poor star sample mentioned above (180 stars). The second peak of the distribution at $V_{0}\sim 15$ appears reflecting the sample selected to cover extremely metal-poor ranges.

\section{High-resolution spectroscopy}\label{sec:hrs}

\subsection{Observations}

High-resolution spectroscopy for selected candidates of very
metal-poor stars has been obtained with the High Dispersion
Spectrograph \citep[HDS: ][]{Noguchi2002PASJ} of the Subaru Telescope. The spectra cover the wavelength region
4030--6800~{\AA} with $R=36,000$ using the $2\times 2$ CCD binning. 

The list of observing runs is given in Table~\ref{tab:obsrun}.  The observing runs were
scheduled in units of half a night. The third column of the table
gives the total number of nights for each run. The number of objects
observed in each run is provided in the fourth column. There is
duplication of objects that were observed in more than one observing
runs.

%The observations were conducted by three {\it Normal Programs} and one {\it
%  Intensive Program} of the Subaru Telescope.

We note that a portion of observing time of the {\it intensive program} was
used for samples of other programs (e.g., stars in moving groups, Li-rich
stars). One night of the August 2017 run was applied to observing candidates of extremely metal-poor stars with
the HDS setup for the blue range. 

\subsection{Data reduction and data quality}

Data reduction was carried out using the standard reduction procedure of the IRAF
echelle package for HDS. It includes CCD bias level subtraction and linearity correction \citep{Tajitsu2010PNAOJ}, cosmic-ray removal following the procedure of \citet{Aoki2005ApJ}, scattered-light subtraction and flat-fielding for 2D images. The spectra were extracted from the images and wavelengths were assigned using Th-Ar calibration spectra. The uncertainty of the wavelength calibration is less than 0.01~{\AA} for our $R=36,000$ spectra.  Sky background light was subtracted in the process of extraction when it is not negligible, i.e., more than a few percent of the peak count of the object. The observations have been conducted avoiding the targets close to the moon; however, the contamination of background light is significant in bright nights when the sky is covered by thin clouds. The process of sky subtraction in the data reduction was examined by visual inspection of the spectra for wide absorption features in the solar spectrum, e.g., the \ion{Mg}{1} b lines at around 5180~{\AA}, and the \ion{Fe}{1} line at 4380~{\AA}. A few spectra that are particularly affected by background light show unrealistically deep absorption lines in the blue regions, such as H$\gamma$ and H$\delta$ most likely due to over-subtraction of the sky spectrum. Such spectra are excluded from the sample for the abundance analysis in this work. The effect of sky is relatively severe in the data obtained in a night in 2016 May.

\begin{figure}
\plotone{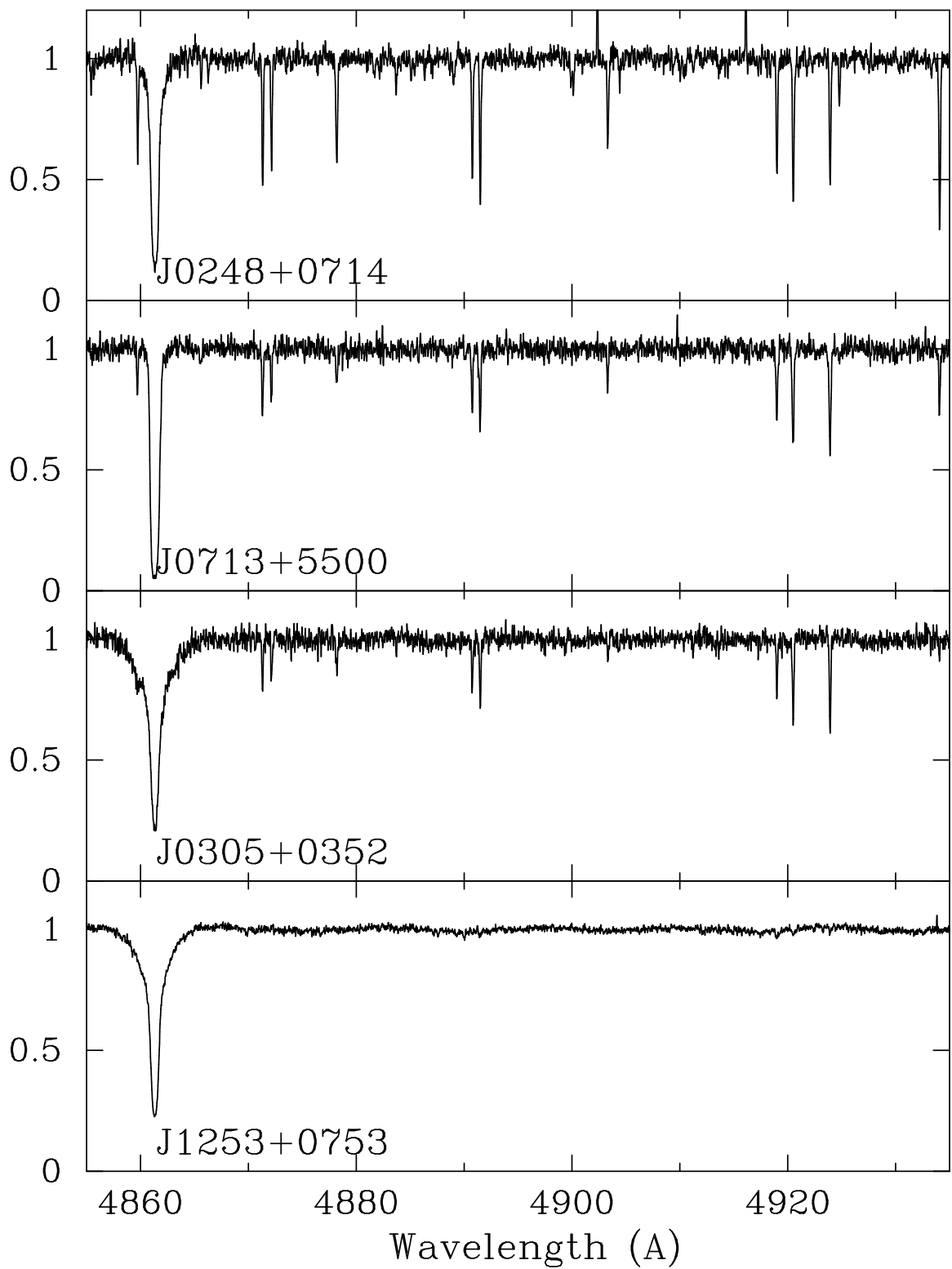}
 \caption{Examples of spectra of a metal-poor giant, an extremely metal-poor giant, a metal-poor turn-off star, and an extremely metal-poor turn-off star from top to bottom. The name of object is presented in each panel.}\label{fig:sp4900}
\end{figure}

\begin{figure}
% \begin{center}
%  \includegraphics[width=10cm]{sp5200.eps}%{c1v.eps}
% \end{center}
\plotone{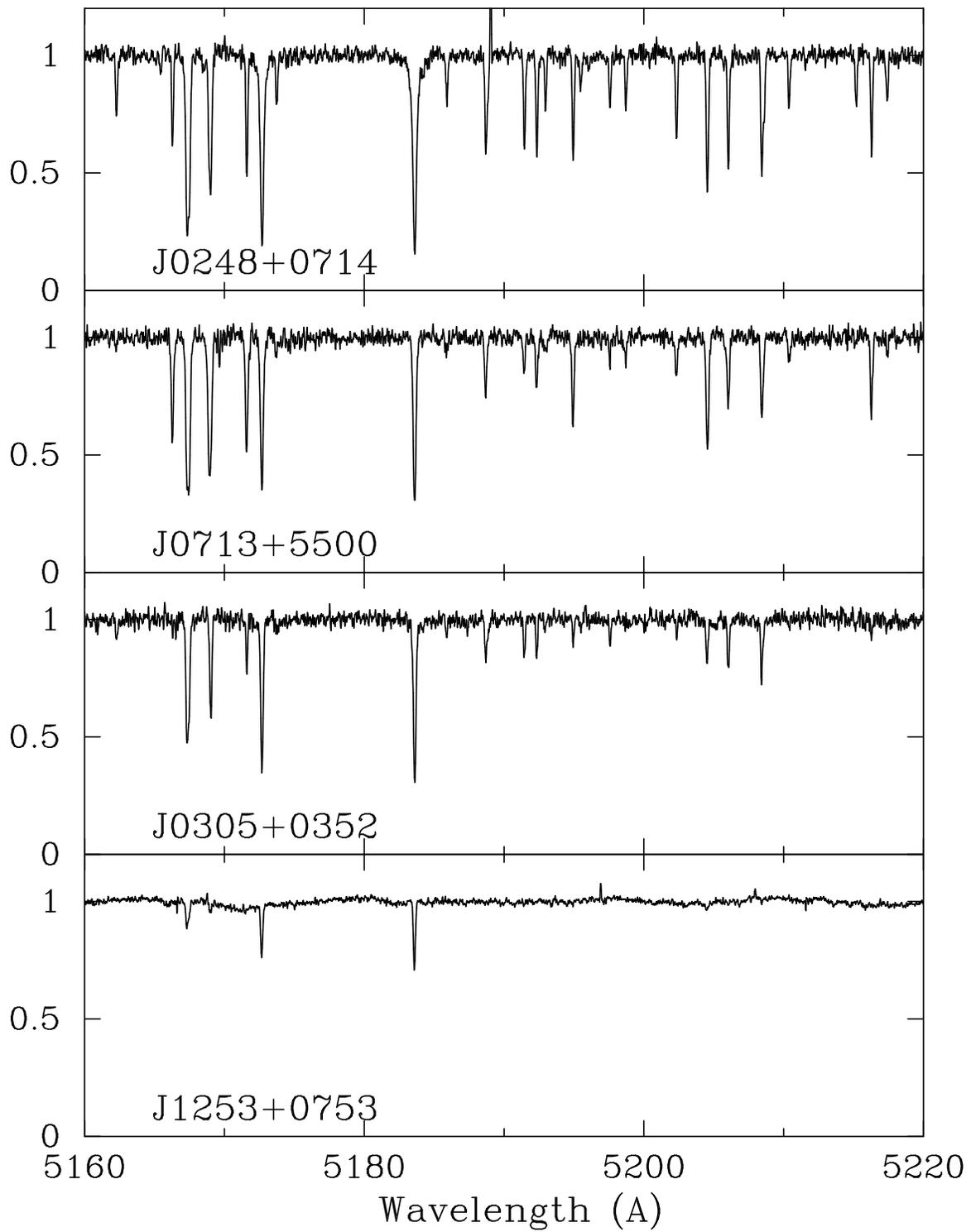}
 \caption{The same as Figure~\ref{fig:sp4900}, but for a different wavelength range.}\label{fig:sp5200}
\end{figure}

When there were multiple exposures for a star, the spectra obtained for individual exposures were combined for each object by summing up the individual exposures. The spectra extracted for individual echelle orders are combined by summing counts for overlapping wavelength ranges between adjacent orders. To obtain a normalized spectrum, a fit of the continuum level of spectra for individual echelle orders is made, and the derived profiles of the continuum are combined by summing counts as done for the original spectra. This provides a combined continuum spectrum. A normalized spectrum for the whole wavelength range is obtained by dividing the combined spectrum by the combined continuum spectrum. Examples of two portions of spectra are presented in Figures~\ref{fig:sp4900} and \ref{fig:sp5200}.

\begin{figure}
\plotone{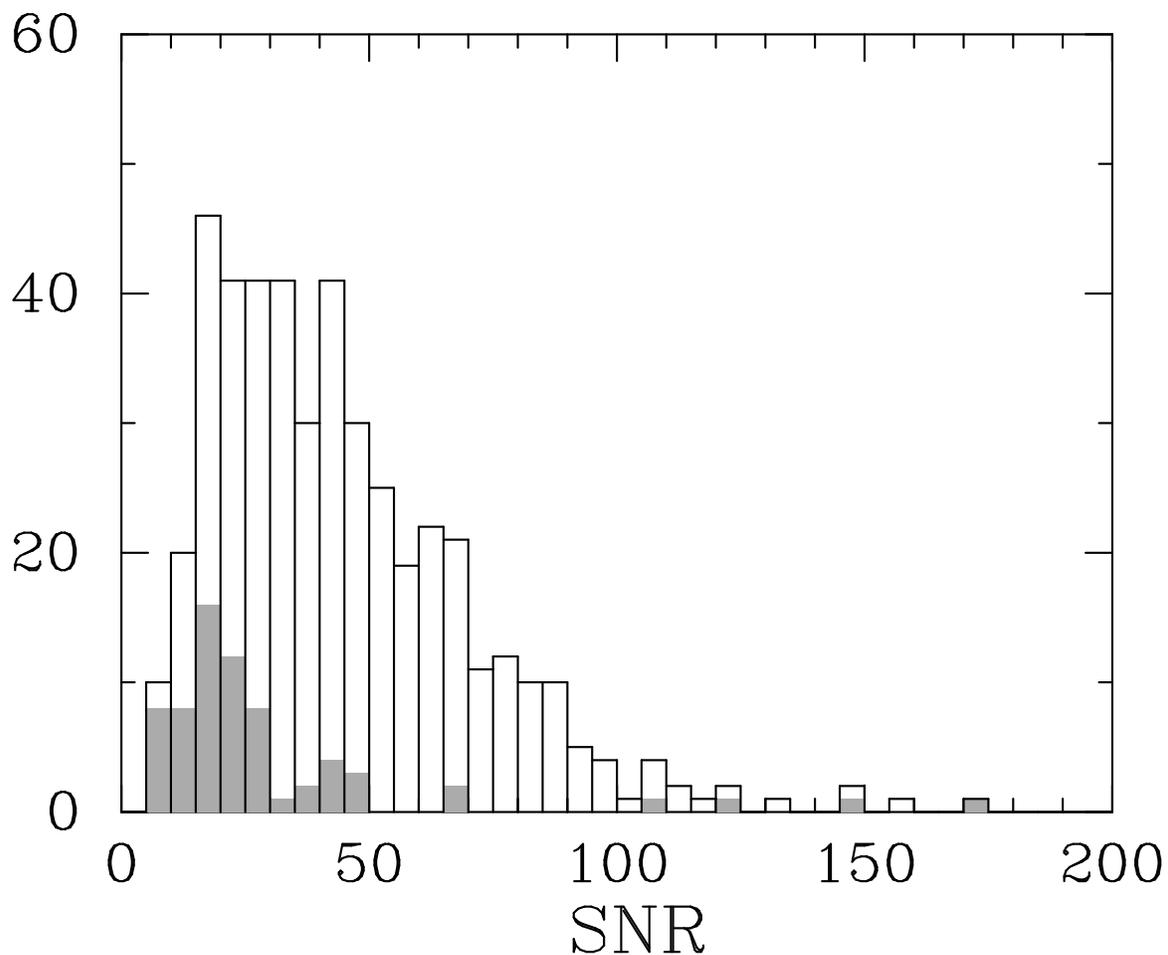}
%\plotone{rvcomp2.eps}
 \caption{Distribution of the SNR (per pixel) of the whole sample in the program. The data not used for abundance analysis due to insufficient SNR or duplication of exposures are shown by the gray histogram. The object with the highest SNR is a double-lined spectroscopic binary (J~0241+0946), which is not included in the abundance study in Paper II.}\label{fig:snr}
\end{figure}

The quality of the spectra varies considerably, because we set
the minimum exposure time to 10~minutes, by which quite high SNR is
achieved for the brightest targets, whereas only low SNR ($\sim 15$) is achieved 
by much longer exposures ($\sim$ 30~minute) for the faintest ones. The SNR is also dependent on the observing condition. 
Figure~\ref{fig:snr} shows the distribution of the SNR of data studied in this paper for radial velocity measurements.  The SNR is estimated from the standard deviation of the data count in the wavelength range almost free from stellar spectral lines around 4500~{\AA}. The value given in Table~\ref{tab:obs}, and presented in the figure, is the SNR per pixel (1.8~km~s$^{-1}$). The SNR per resolution element is approximately 2.2 times larger than that in the table and the figure. The SNR is also well approximated by the square root of the photon counts (Table~\ref{tab:obs}), however, it is significantly lower in some objects that are affected by sky background. 

The sample consists of 445 spectra of 420 stars including duplicated exposures.
 The radial velocities are investigated for the whole sample.
The final sample of the chemical abundance measurements  (385 stars) was determined through the abundance analysis process as the data quality is sufficiently high for
the purpose taking account of the SNR and the number of Fe lines that can be used for abundance measurements, which depends on the effective temperature and metallicity (see Paper II  for the details). The distribution of the SNR for the data not used for the abundance analysis is overplotted as a gray histogram in the figure. 

\section{Radial velocity measurements}\label{sec:rv}

Radial velocities of the program stars are determined through cross-correlation 
with the Subaru high-resolution spectrum for a VMP standard star, using the IRAF/{\it fxcor}. 
Fifteen orders in the wavelength range from 4230 through 5210~{\AA} are used for each spectrum for the cross-correlation.  
The average value and the standard deviation of the 15 measured radial velocities 
are regarded as the adopted radial velocity and measurement uncertainty, respectively, for the corresponding spectrum. 
The typical uncertainty of the adopted radial velocity is about 0.14\,km\,s$^{-1}$. Including the systematic errors due to instability of the instrument mostly due to temperature changes, the errors of radial velocity measurements from the Subaru spectra is $\lesssim 1$~km~s$^{-1}$.  
The radial velocities measured from high-resolution spectra are given in
Table~\ref{tab:obs}. 

\begin{figure}
\includegraphics[width=12cm]{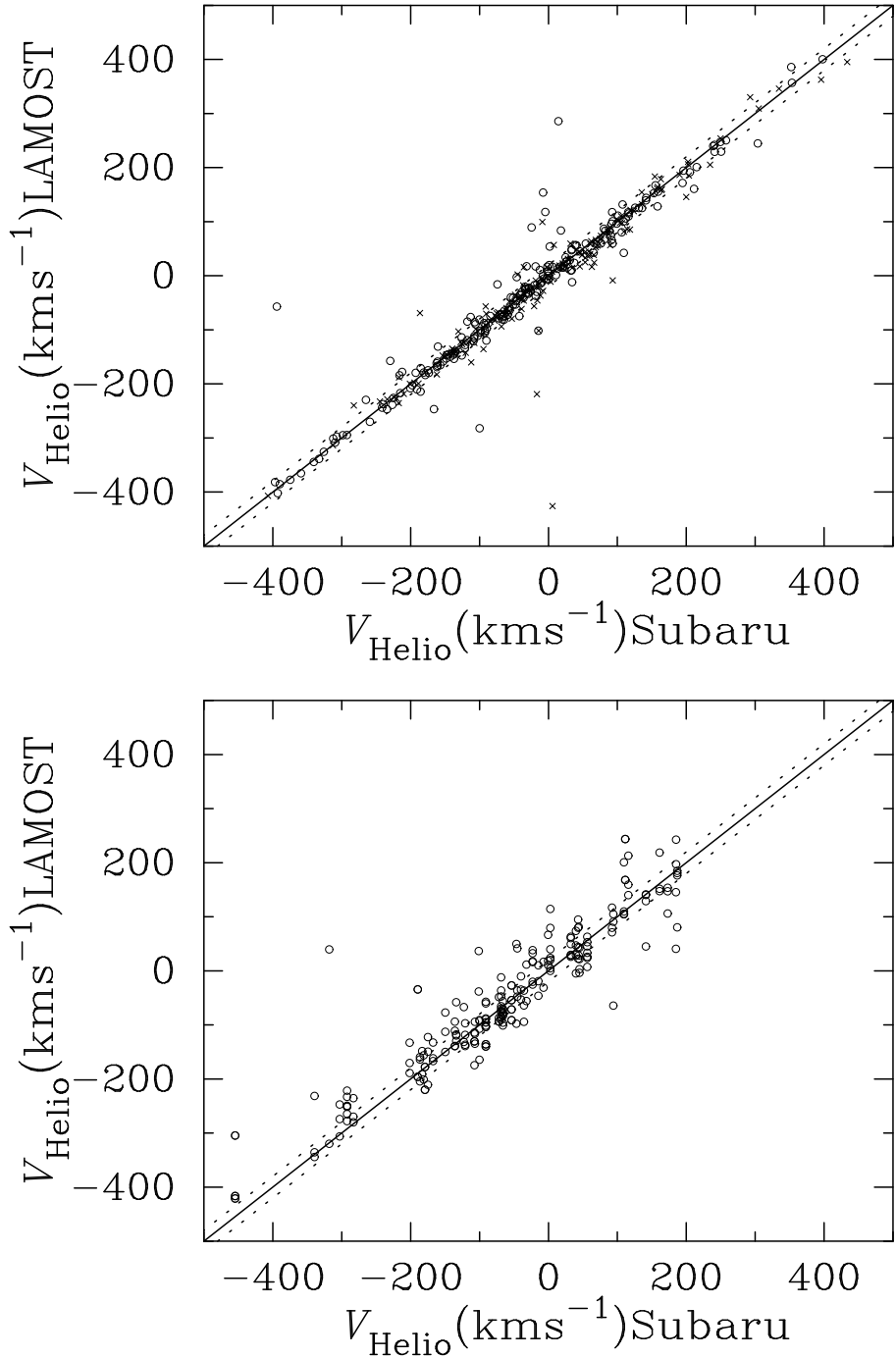}
%\plotone{rvcomp1.eps}
%\plotone{rvcomp2.eps}
 \caption{Comparison of the heliocentric radial velocities obtained from the Subaru high-resolution spectra with LAMOST measurements. Upper panel: Objects for which no variation of radial velocity is detected by LAMOST measurements, including those with single-epoch LAMOST observation. Lower panel: Objects for which radial velocity variation ($>20$~km~s$^{-1}$) is detected in the LAMOST measurements. The solid lines indicate one-to-one relationship, and the dotted lines are those shifted by 20~km~s$^{-1}$.}\label{fig:rvcomp}
\end{figure}

Comparisons with radial velocities obtained from the LAMOST data
release (DR5) are shown in Figure~\ref{fig:rvcomp}. No useful radial velocities are available from DR5 for several stars, some of which are most metal-poor objects. 
The sample is separated into two subsamples. One is a group of stars that show
variations of radial velocities by more than 20~km~s$^{-1}$ in the
multi-epoch measurements with LAMOST.  The criterion (20~km~s$^{-1}$) is adopted taking account of the errors of the radial velocity measurements from LAMOST low-resolution spectra for our metal-poor stars (5-20~km~s$^{-1}$) in DR5. The subsample consists of 56 stars. Variations of radial velocities
suggest that the objects belong to binary systems with companions of faint main-sequence stars or white dwarfs, in the case of
metal-poor stars around the main-sequence turn-off or the red giant
branch. These stars are shown in the lower panel of
Figure~\ref{fig:rvcomp}  with all the radial velocity data taken from LAMOST DR5.  Significant variations of the radial
velocities from LAMOST are found due to the above selection. Nevertheless, there is still 
a clear correlation between the values from LAMOST and Subaru.
The mean and root-mean-square (r.m.s.) of the difference between radial velocities determined with Subaru and LAMOST ($V_{\rm Helio}$(Subaru) $-V_{\rm Helio}$(LAMOST)) are +3.6 and 14.0~km~s$^{-1}$, respectively, excluding the data with differences larger than 40~km~s$^{-1}$, i.e. twice the above criterion, which should be clearly affected by binary motions.  

The other subsample includes stars with single-epoch LAMOST mearurements (201 stars), and
those showing no significant variation of radial velocities
($<20$~km~s$^{-1}$) in the LAMOST measurements (154 stars). The overall agreement
between the radial velocities from LAMOST and Subaru is fairly
good.  The mean and r.m.s. of the difference between radial velocities determined with Subaru and LAMOST ($V_{\rm Helio}$(Subaru) $-V_{\rm Helio}$(LAMOST)) are +2.4 and 13.9~km~s$^{-1}$, respectively, excluding the data with difference larger than 40~km~s$^{-1}$.
The difference of the radial velocities obtained by high-resolution spectroscopy from the LAMOST measurements is less than 20~km~s$^{-1}$ for 158 objects among the 201 stars with single-epoch LAMOST observations. For 112 stars with multi-epoch LAMOST observations, the difference between the measurements of high-resolution spectroscopy and the average of the radial velocity of LAMOST measurements is less than 20~km~s$^{-1}$. Hence, there is no signature of binarity for these 270 stars at this moment. 

The radial  velocities measured from high-resolution spectra of 85 stars  differ from the measurements based on LAMOST spectra  by more than 20~km~s$^{-1}$.
This set of stars consists of 43 stars with single-epoch and 42 with multiple-epoch LAMOST observations.
These stars, as well as the 56 stars with large scatter of the radial velocities from the LAMOST measurements, are candidates of binary stars. 
Follow-up observations
of these objects are desired to confirm their binarity and determine their orbital parameters. We note that five of the seven double-lined spectroscopic binaries discussed in \S~\ref{sec:sb2} are included in the sample of objects that show radial velocity variations. 

%Radial velocities obtained by high-resolution spectroscopy show significant departure from the LAMOST measurements for the remaining 86 stars, i.e., 43 stars with single-epoch LAMOST observations and the other 43 stars with multi-epoch LAMOST ones. 

In addition to the spectroscopic information, the astrometry from the Gaia mission can also be used to constrain the binarity of the objects.
Gaia's renormalized unit weight error (RUWE) is known to be sensitive to the presence of unresolved companions \citep{Belokurov2020MNRAS,Stassun2021ApJ}. In our sample, there are 21 objects having RUWE$>1.4$. These objects are likely to have companions. Among them, seven stars show radial velocity variations, including J~0119-0121 and J~1435+1213, that show large variations among LAMOST measurements. For the remaining objects, no variation of radial velocity (larger than 20~km~s$^{-1}$) is found. The frequency of stars that show radial velocity variations is 7/21 (33\%), which is similar to the frequency in the whole sample (34\%). The no excess of the frequency among the objects with large RUWE would at least be partially because the number of  radial velocity measurements with LAMOST and Subaru is still quite limited. Another possible reason is that large RUWE could be found for objects with near face-on view, for which radial velocity variation is not detectable.

\begin{figure}
\plotone{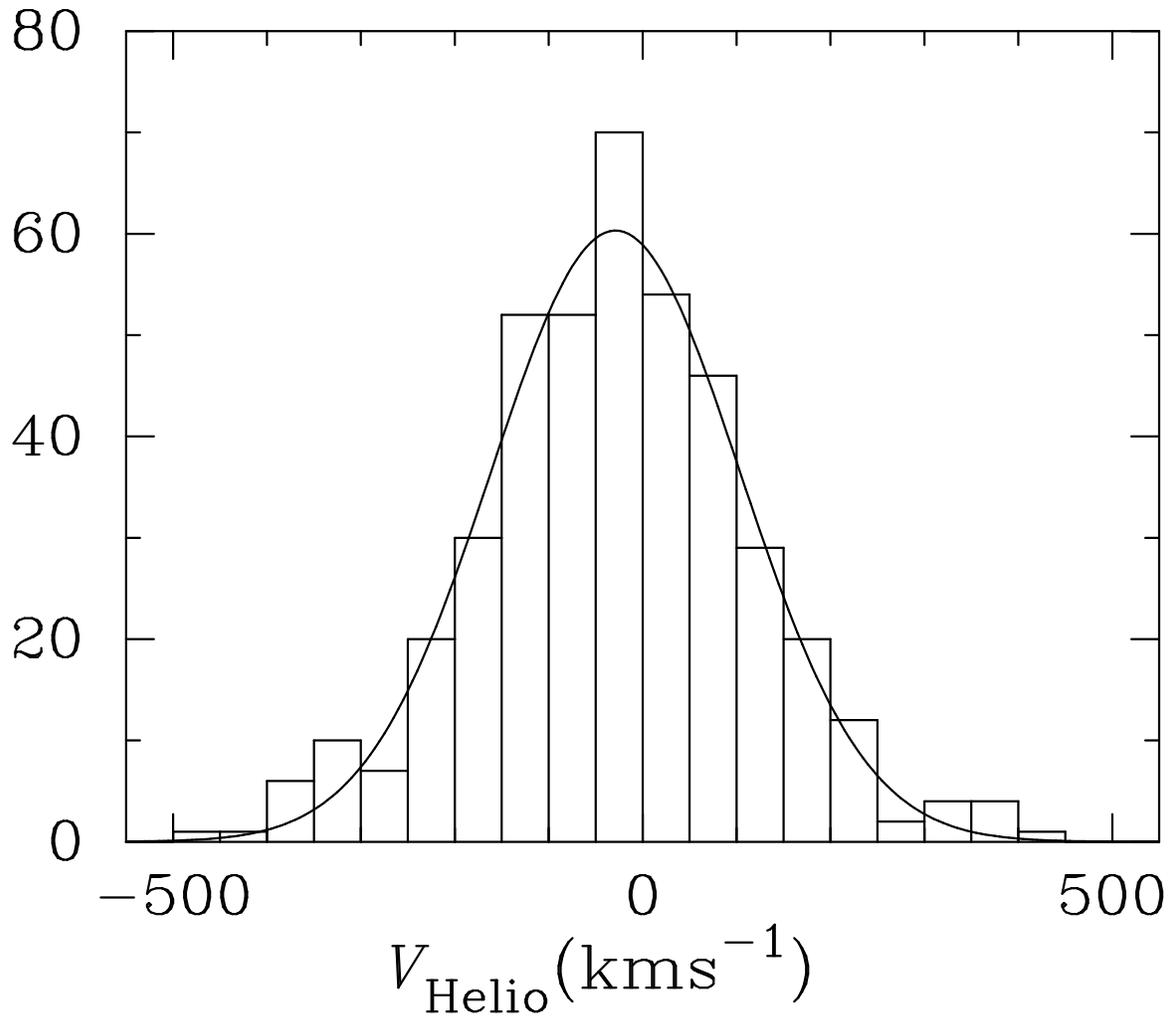}
 \caption{Distribution of the heliocentric radial velocities of our sample.}\label{fig:rvhist}
\end{figure}

Figure~\ref{fig:rvhist} shows the distribution of heliocentric radial velocities of our sample. The distribution is well fit by a Gaussian with a central value of $V_{\rm Helio} = -29$~km~s$^{-1}$. The offset of the central value is likely  due to the selection bias of our sample that includes larger number of objects in the northern sky. 

\section{Interstellar absorption}

We measured the equivalent widths of interstellar \ion{Na}{1} D absorption
lines when the interstellar component is separated from the stellar
absorption. The measurements are made by direct integration of the interstellar absorption of D$_{1}$ and D$_{2}$ lines separately by IRAF/{\it splot}. The equivalent widths of both lines are measured for 324 objects. 

\begin{figure}
\plotone{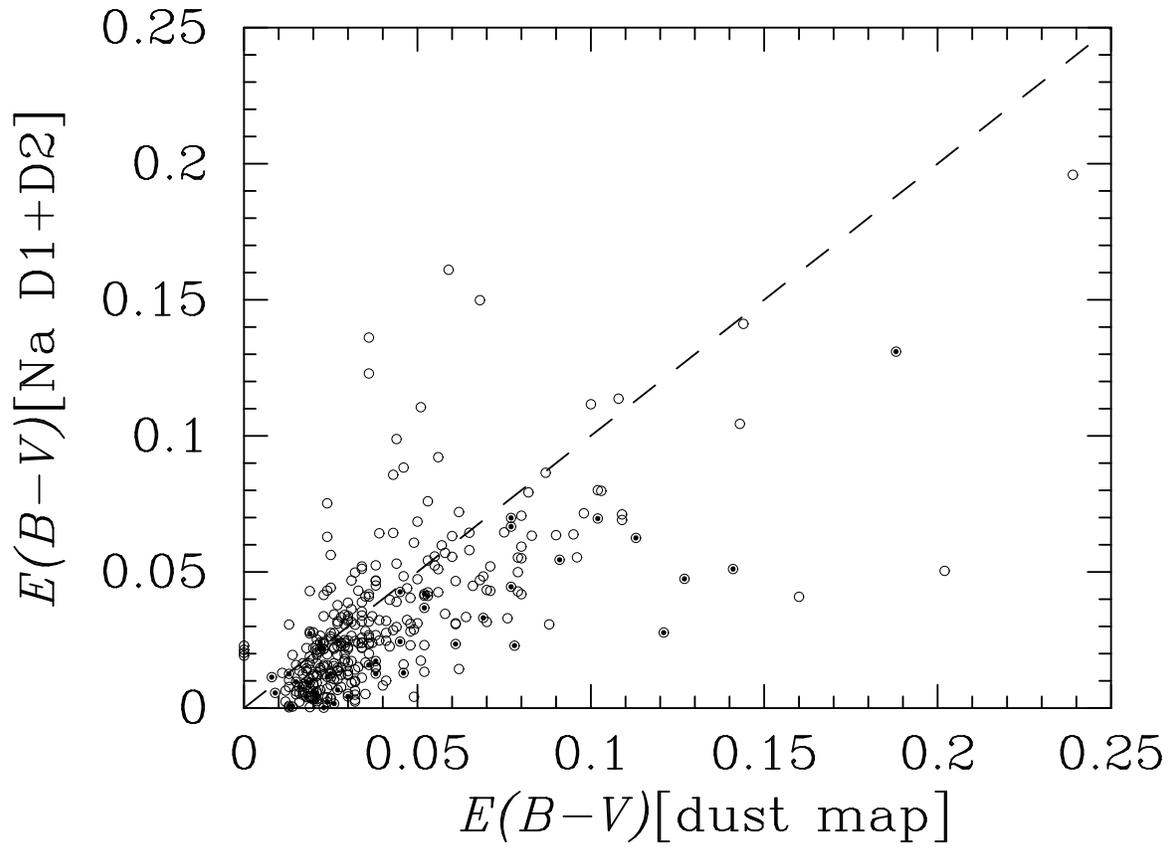}
 \caption{Reddening obtained from the Na D$_{1}$ and D$_{2}$ line absorption compared with that from the dust map. Filled circles are objects with distances smaller than 0.5~kpc.}\label{fig:ebv}
\end{figure}

We estimate the interstellar reddening, $E(B-V)$, adopting the relation between the reddening and the Na D line absorption given by \citet{Poznanski2012MNRAS}. We adopt the relation given by their equation (9) for the equivalent width $W$ combined for D$_{1}$ and D$_{2}$ lines. Their formula has a lower limit of $\log E(B-V)=-1.85$ ($E(B-V)=0.014$). For the weakest lines ($W<0.15$~{\AA}) we assume a linear relation between $W$ and $E(B-V)$ as it connects to the above relation for larger $W$. Figure~\ref{fig:ebv} shows a comparison of the $E(B-V)$ values obtained from the Na D line measurement with those derived from a dust map \citep{Green2018MNRAS}. A fairly good correlation is found for most of the stars, whereas larger reddening is derived from the dust map for a dozen of stars. Some of the stars having larger $E(B-V)$ from the dust map than that from the interstellar absorption are nearby objects: This is found in the figure, where objects with a distance smaller than 0.5~kpc are presented by filled circles. For these objects the reddening might be overestimated from the dust map.  On the other hand, there are several stars that have larger $E(B-V)$ from Na D lines than that from the dust map. We confirm by visual inspection of the spectra that the five objects with $E(B-V)$[Na D]$>0.1$ and $E(B-V)$[dust map]$<0.1$ exhibit strong interstellar absorption of Na D lines. The discrepancy of the $E(B-V)$ values for these stars might indicate scatter in the relation between the Na interstellar absorption and the dust extinction.

\section{Double-lined spectroscopic binaries}\label{sec:sb2}

\begin{figure}
% \begin{center}
%  \includegraphics[width=10cm]{sb2_5200.eps}%{c1v.eps}
% \end{center}
\plotone{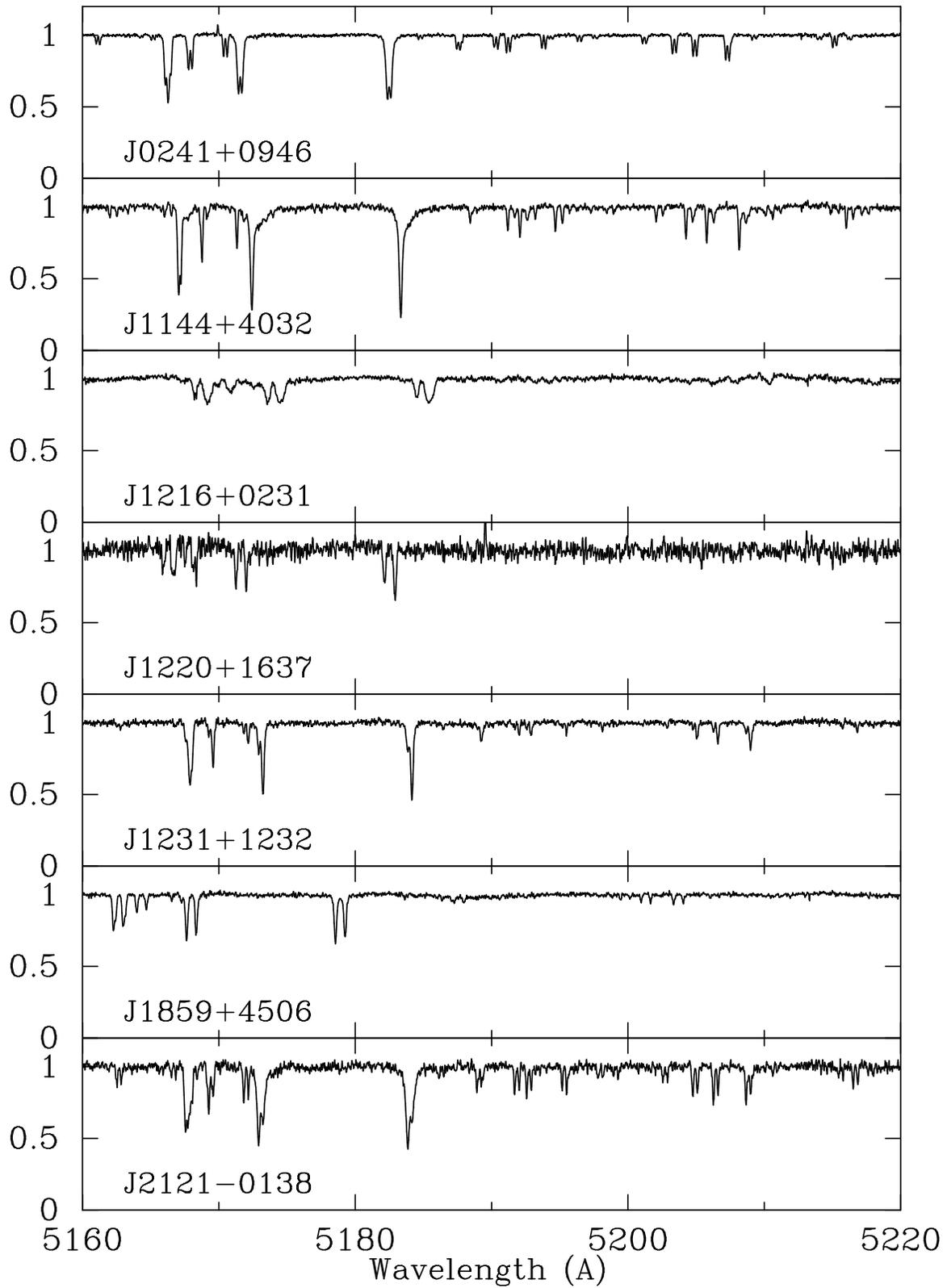}
 \caption{Spectra of double-lined spectroscopic binaries. The names of the objects are presented in the panel.}\label{fig:sb2}
\end{figure}

Among the 420 stars for which high-resolution spectra are obtained, seven stars are identified as double-lined
spectroscopic binaries that clearly exhibit pairs of absorption
lines. Examples of spectral features are shown in Figure~\ref{fig:sb2}. 

The radial velocity of each component is measured for Fe {\small I}
lines that are isolated from other spectral features. Measurements are
carried out by fitting Voigt profiles to two components separately using IRAF
{\it splot}. The results are given in Table~\ref{tab:sb2}. The number of lines
used in the analysis depends on the strengths of the
absorption features and the data quality. 

We also applied cross-correlation between the spectra of these stars and a template spectrum of a single star. Double peaks of the cross-correlation are found for stars with double lines that are clearly separated (e.g., J1220+1637, J1859+4506; see Figure~\ref{fig:sb2}), confirming the radial velocities determined from Fe {\small I} lines.
We find that the second peak is not clear for stars that have only weak component of the secondary as expected. The radial velocities from Fe {\small I} lines are adopted as the final result.

Absorption features of the stronger component in the J~1216+0231 spectrum are
slightly broader than those of the other component, as well as broader than the absorption
features of the other stars. This might indicate that it is a triplet system, as SDSS~J1108+1747 reported by \citet{Aoki2015AJ}. Further observations of this object at different phases will be needed to understand this system. 

The velocity differences between the two components are 15 -- 54
km~s$^{-1}$. This range overlaps with that of the three double-lined spectroscopic binaries studied
for the metal-poor stars found with SDSS and follow-up high-resolution spectroscopy \citep{Aoki2015AJ}. The period
and the binary separation of these systems would be smaller than
1000~days and several AU, respectively. 

\begin{figure}
% \begin{center}
%  \includegraphics[width=11cm]{sb2cmd.eps}%{c1v.eps}
% \end{center}
\plotone{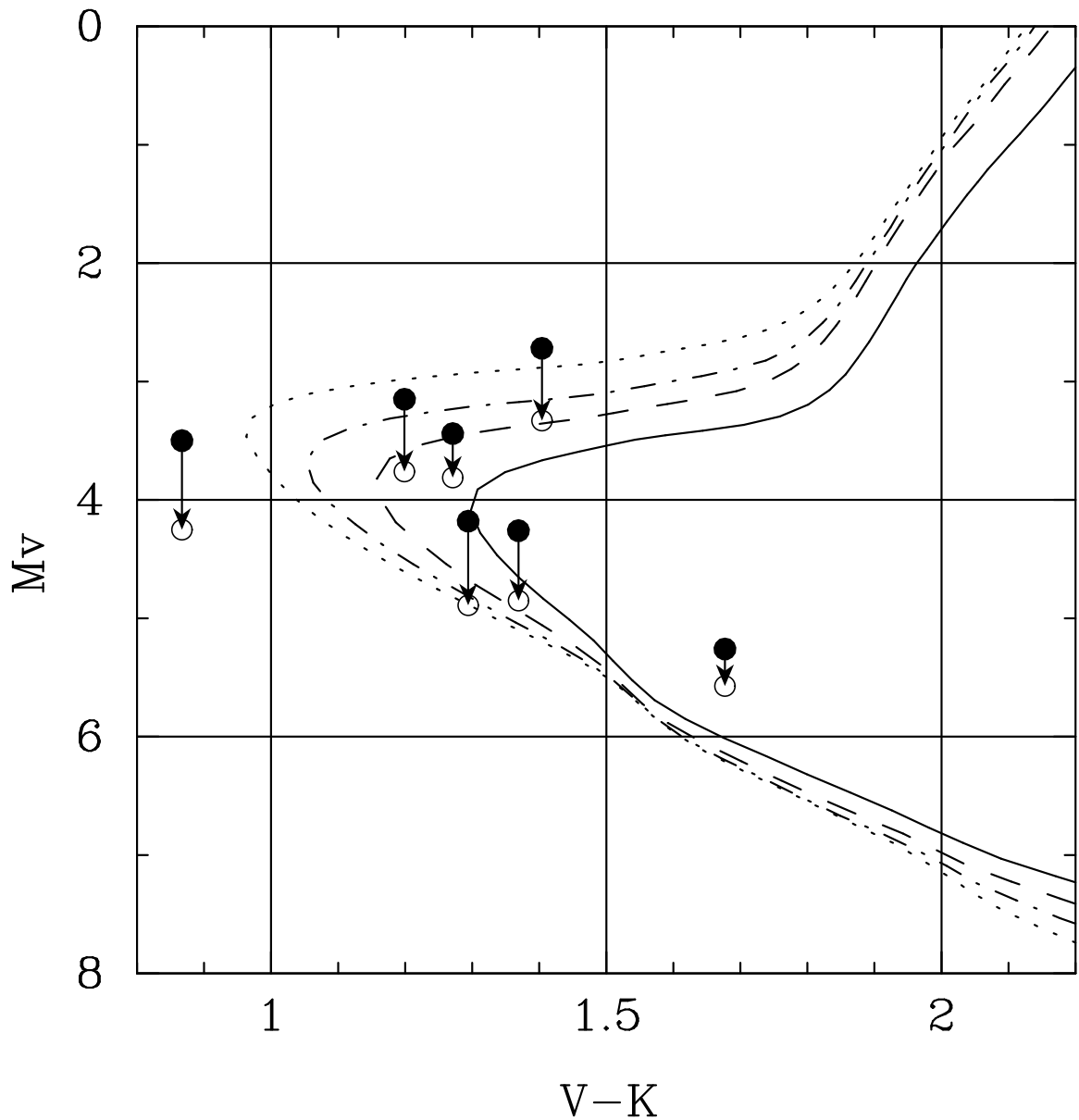}
 \caption{Color-Magnitude diagam for the double-lined spectroscopic binaries. The $(V-K)_{0}$ and $M_{V}$ for the objects are shown by filled circles. The open circles indicate the $M_{V}$ values estimated for the primary stars in the systems (see text). The lines show the isochrones of Yonsei-Yale models \citep{Kim2002ApJS} with enhancement of $\alpha$ elements for 12~Gyr for [Fe/H]$=-1.5$ (solid), $-2.0$ (dashed), $-2.5$ (dot-dashed), $-3.5$ (dotted).}\label{fig:cmd}
\end{figure}

The $(V-K)_{0}$ color and absolute $V$ magnitude ($M_{V}$) of the objects are
shown in Figure~\ref{fig:cmd}. The absolute magnitude is derived from
the $V_{0}$ magnitude and the distance estimated by the parallax of {\it
  Gaia} DR2 \citep{Gaia2016AA, Gaia2018AA}. Isochrones of the Yonsei-Yale stellar evolution models 
\citep{Kim2002ApJS} are shown for the metallicity range $-3.5\leq$[Fe/H]$\leq -1.5$. The
double-lined spectroscopic binaries found by our study are located in
the main-sequence turn-off range. This is expected because
the luminosity is very sensitive to mass in red giants and, hence, the
primary is dominant in the system if it has already evolved to a red
giant.

We here estimate the contribution of each component of the binary
system to the overall flux from the strengths of the absorption lines. The depth of the
absorption lines of Mg I b lines around 5180~{\AA} is not very sensitive to
the effective temperature in the range of the main-sequence
turn-off ($5500$~K$<$\tefft$<6500$ ~K). Hence, we assume that the apparent depths of the Mg lines are approximately proportional 
to the contributions of the two components to the flux in the $V$ band. The open
circles in Figure~\ref{fig:cmd} indicate the absolute magnitudes of the
primaries. Here the color of the primary is assumed to be the same as
that of the system. This is a good approximation for systems in
which the primary is dominant, or when the two components have similar
luminosity and \tefft. The most uncertain case is a system that includes a subgiant, since the luminosity (absolute magnitude) is similar in a relatively wide ranges of {\tefft} and $V-K$ color. The 10\% uncertainty in the estimate of the contribution of a component to the luminosity results in about $\pm 250$~K uncertainty of {\tefft} according to the isochrone. 
%{\it The primary's $(V-K)_{0}$ could be smaller in the
%intermediate cases at most xx magnitude.}

Figure~\ref{fig:cmd} clearly indicates that the primary stars are on the 
main-sequence in most cases. The secondary should also be on the 
main-sequence, having larger $(V-K)_{0}$ color, with similar or larger
(fainter) absolute magnitude, than the primary. An exception is J~1216+0231
($M_{V}=2.7$) whose primary star would be on the subgiant branch.  This system could be triplet as mentioned above. To understand the system, further monitoring of the variations of spectral features is desirable.

We estimate the metallicity of the primary stars adopting the equivalent
widths of the Fe lines used to measure the radial velocities. The
equivalent widths are corrected by the contribution of the primary
star to the flux estimated above (Table~\ref{tab:ewsb2}). The effective temperature is estimated from colors
as done for our whole sample that will be reported in Paper
II. The surface gravity and microturbulent velocity are fixed to $\log
g =4.0$ and $v_{\rm t}=1.5$~km~s$^{-1}$, respectively, which are sufficient for the
present purpose. The metallicity ([Fe/H]) of our sample ranges from $-1.8$ to
$-3.2$, confirming that they are very metal-poor, although the uncertainty of
the [Fe/H] values is as large as 0.5~dex, primarily due to the uncertainties in
the estimate of the contribution of the primary star to the flux.

The above analysis indicates that the seven objects are all binary
systems of very metal-poor stars with short periods ($<1000$~days), as
the three objects studied by \citet{Aoki2015AJ} from the SDSS sample. The
number of main-sequence turn-off stars ($T_{\rm eff} >$ 5500~K) in our
whole sample is 242. The fraction of double-lined spectroscopic
binaries detected by single-epoch observations, 7/242= 2.9\%, is similar to
that obtained by \citet{Aoki2015AJ}: 3/109=2.8\%. We note that the distribution
of effective temperature in $T_{\rm eff}>$ 5500~K of the whole sample
is also similar to that of \citet{Aoki2015AJ}. \citet{Aoki2015AJ} discussed
the frequency of binary systems with short periods at very low
metallicity is at least 10~\%, taking the detection
probability of the systems by a single epoch observation into account. The result of
the present study supports this argument by increasing the sample
size. 

The frequency of binaries has been studied by a variety of spectroscopic survey projects. \citet{Moe2019ApJ} recently investigated the metallicity dependence of binary frequency by compiling previous studies including the Carney-Lathum sample \citep[e.g., ][]{Carney2005AJ,Latham2002AJ}, \citet{Rastegaev2010AJ}, SDSS and LAMOST \citep{Hettinger2015ApJ, Gao2017MNRAS}, SDSS/APOGEE \citep{Badenes2018ApJ}, and very metal-poor stars \citep{Hansen2015A&A}. Applying corrections for completeness of the survey depending on metallicity, i.e. lower completeness for metal-poor stars due to weakness of absorption lines, they conclude that the fraction of close binaries ($P<10,000$ days) is higher at lower metallicity, reaching about 50\% at [Fe/H]$=-3$. It should be noted that the original sample of very metal-poor stars referred to in their work is still small: 91 stars with [Fe/H]$<-0.9$ were studied by \citet{Carney2005AJ} and 41 r-process-enhanced stars with [Fe/H]$<-1.6$ were studied by \cite{Hansen2015A&A}. Although the frequency of binaries directly estimated for these samples is 15-20\%, the correction estimated by \citet{Moe2019ApJ} is as large as a factor of three at the lowest metallicity. The frequency of double-lined spectroscopic binaries with $P<1000$ days estimated by our study is well below the frequency of all binary systems that include pairs of stars having significantly different luminosity. 

At higher metallicity ([Fe/H]$>-1$), a much larger number of double-lined spectroscopic binaries has been identified by high-resolution spectroscopic surveys. \citet{ElBadry2018MNRAS} reported the detection of about 2500 binaries from 20,000 main-sequence stars of the APOGEE sample. More recently, \citet{Traven2020AA} studied more than 580,000 stars of the GALAH survey sample (including red giants), identifying 12,760 double-lined spectroscopic binaries. Most of them are main-sequence stars as expected. The fraction of double-lined spectroscopic binaries in their sample, about 2\%, should be significantly higher if the sample is limited to main-sequence. However, the frequency estimated by the present work for VMP/EMP stars, about 10\%, seems to be higher than that found for metal-rich stars, supporting theoretical studies of binary formation at very low metallicity \citep[e.g., ][]{Machida2008ApJ}.

\section{Summary}

We acquired high-resolution spectroscopy using the Subaru Telescope High Dispersion Spectrograph for about 400 very metal-poor stars selected by means of  LAMOST low-resolution spectroscopy. This paper reports on the observing program, including sample selection, observations and data reduction, estimates of extinction from interstellar \ion{Na}{1} D lines, and radial velocities. Based on the detection of seven double-lined spectroscopic binaries, the frequency of binary systems with short periods is discussed. The data are used for detailed abundance analyses that are reported separately in paper II of this series. 

\acknowledgments

This research is based on data collected at Subaru Telescope, which is operated by the National Astronomical Observatory of Japan. We are honored and grateful for the opportunity of observing the Universe from Maunakea, which has the cultural, historical and natural significance in Hawaii.
Guoshoujing Telescope (the Large Sky Area Multi-Object Fiber
Spectroscopic Telescope, LAMOST) is a National Major Scientific
Project built by the Chinese Academy of Sciences.  Funding for the
project has been provided by the National Development and Reform
Commission.  LAMOST is operated and managed by the National
Astronomical Observatories, Chinese Academy of Sciences.
This work has made use of data from the European Space Agency (ESA) mission
{\it Gaia} (\url{https://www.cosmos.esa.int/gaia}), processed by the {\it Gaia}
Data Processing and Analysis Consortium (DPAC,
\url{https://www.cosmos.esa.int/web/gaia/dpac/consortium}). Funding for the DPAC
has been provided by national institutions, in particular the institutions
participating in the {\it Gaia} Multilateral Agreement.
This work was supported by JSPS - CAS Joint Research Program.  WA and
TS were supported by JSPS KAKENHI Grant Numbers 16H02168, 16K05287 and
15HP7004. MNI was supported by JSPS JSPS KAKENHI Grant Numbers 17K14249 and 20H05855. 
The Chinese team was supported by NSFC Grant Nos. 11988101, 11973049, 11890694, 11625313, 11973048, 11927804, 12090044, 11833006, 12022304, 11973052, 
National Key R\&D Program of China No.2019YFA0405502, the Strategic Priority Research Program of Chinese Academy of Sciences, Grant No. XDB34020205, 
the science research grants from the China Manned Space Project with No. CMSCSST-2021-B05, 
and the Youth Innovation Promotion Association of the CAS (id. Y202017, 2019060, 2020058). 
N.C. acknowledges funding by the Deutsche Forschungsgemeinschaft (DFG, German Research Foundation) -- Project-ID 138713538 -- 
SFB 881 ("The Milky Way System", subproject A04).

%HL was supported by NSFC grants No. 11573032, 11390371.


\vspace{5mm}
\facilities{LAMOST, The Subaru Telescope}


%\software{astropy \citep{2013A&A...558A..33A},  
%          Cloudy \citep{2013RMxAA..49..137F}, 
%          SExtractor \citep{1996A&AS..117..393B}
%          }



\clearpage

%\begin{figure}
% \caption{Distribution of the $S/N$ ratios of the %sample.}\label{fig:sn}
%\end{figure}

%\begin{figure}
% \includegraphics[width=12cm]{rvcomp2.eps}
%%\plotone{rvcomp2.eps}
%%\plotone{rvcomp2.ps}
% \caption{The same as Fig.\ref{fig:rvcomp2}, but for velocity differences %($\Delta V_{\rm Helio}=V_{\rm LAMOST}-V_{Subaru}$).}
%\label{fig:rvcomp2}
%\end{figure}

\clearpage

\begin{deluxetable*}{llccccccccc}
%\tablenum{1}
\tablecaption{Objects studied with high-resolution spectra\label{tab:obj}}
\tablewidth{0pt}
\tablehead{
\colhead{Short ID} & \colhead{Object name} & \colhead{$V$} & \multicolumn{2}{c}{$E(B-V)$} &\colhead{$B_{0}$} &\colhead{$V_{0}$} &\colhead{$g_{0}$}&\colhead{$r_{0}$} &\colhead{$J_{0}$} &\colhead{$K_{0}$} \\
 & & & dust map & Na D & &&&&&  
}
%\decimalcolnumbers
\startdata
J0002+0343 & LAMOST J000235.04+034337.5 & 15.094 &  0.028 &  0.024 & 15.446 & 15.019 & 15.068 & 14.894 & 14.062 & 13.738 \\
J0003+1556 & LAMOST J000310.89+155601.8 & 12.321 &  0.058 &  0.035 & 12.747 & 12.166 & 12.368 & 11.912 & 10.779 & 10.312 \\
J0006+0123 & LAMOST J000637.82+012343.3 & 13.123 &  0.036 &  0.041 & 13.793 & 13.028 & 13.487 & 13.008 & 11.856 & 11.470 \\
J0006+1057 & LAMOST J000617.20+105741.8 & 12.912 &  0.070 &  0.043 & 13.555 & 12.725 & 13.076 & 12.590 & 10.954 & 10.330 \\
J0013+2350 & LAMOST J001351.30+235049.1 & 12.328 &  0.053 &  0.041 & 12.608 & 12.187 & 12.305 & 12.053 & 11.173 & 10.876 \\
\enddata
\end{deluxetable*}

\begin{deluxetable*}{llcc}
%\tablenum{1}
\tablecaption{Observing Runs\label{tab:obsrun}}
\tablewidth{0pt}
\tablehead{
\colhead{Proposal ID} & \colhead{Obs. dates} & \colhead{number of nights} & \colhead{number of objects}  
}
%\decimalcolnumbers
\startdata
S14A-112 & May 9-10,  2014 & 2 & 51  \\
S15A-093 & March 5, 2015 & 1 & 10  \\
S15B-087 & November 28-29, 2015 & 2 & 42  \\
S16A-119I & April 26-27, 2016 & 1 & 18  \\
S16A-119I & May 20, 22, 23, 27, 28, 2016 & 5 & 123  \\
S16A-119I & November 16-19, 2016 & 4$^{a}$ & 49 \\
S16A-119I & February 15-19, 2017 & 5 & 84  \\
S16A-119I & August 1-5, 2017 & 2.5$^{a,b}$ & 27 \\
S16A-119I & January 24-27, 2018 & 2.5$^{c}$ & 0 \\
\enddata
\tablenotetext{a}{Observing time was partially lost by telescope troubles.}
\tablenotetext{b}{Observing time was partially used for the blue setup.}
\tablenotetext{c}{No useful data were obtained by bad weather.}
\end{deluxetable*}

\begin{deluxetable*}{lcrrrrrrr}
%\tablenum{1}
\tablecaption{Observation and kinematics information\label{tab:obs}}
\tablewidth{0pt}
\tablehead{
\colhead{Short ID} & \colhead{date of observation} & \colhead{SNR} & \colhead{$V_{\rm Helio}$} &\colhead{$\sigma(V_{\rm Helio})$} &\colhead{Parallax} & \colhead{error} & \colhead{Distance} & \colhead{error} \\
 & & & \colhead{(km~s$^{-1}$)} & \colhead{(km~s$^{-1}$)} & \colhead{(mas)}
 & \colhead{(mas)} & \colhead{(pc)} & \colhead{(pc)}
}
%\decimalcolnumbers
\startdata
J0002+0343 & 20151129 &    23.7  & $-91.4$ & 0.8 & 0.647 & 0.050 & 1465 & 109 \\
J0003+1556 & 20161116 &    32.2 & $-185.5$ & 0.1 & 0.3623 & 0.0542 & 2380 & 298 \\
J0006+0123 & 20161119 &    44.1 & $-65.9$ & 0.2 & 0.6591 & 0.0368 & 1446 & 78 \\
J0006+1057 & 20161116 &    21.9 & $-312.4$ & 0.3 & 0.1103 & 0.0425 & 4549 & 686 \\
J0013+2350 & 2017 8 4 &  41.4 & $-259.2$ & 0.1 & 2.0484 & 0.0384 & 482 & 9 \\
%J0023+2023 & 20161119 &   2100 & 37 & -126.0 & 0.2 & 0.3983 & 0.0309 & 2295 & 163 \\
\enddata
\end{deluxetable*}

\begin{deluxetable*}{lcccccc}
%\tablenum{1}
\tablecaption{Double-lined spectroscopic binaries\label{tab:sb2}}
\tablewidth{0pt}
\tablehead{
\colhead{Object} & \colhead{$V_{\rm Helio, A}$} & \colhead{error} & \colhead{$V_{\rm Helio, B}$} &\colhead{error} &\colhead{$N$} & [Fe/H] \\
& (km~s$^{-1}$) &(km~s$^{-1}$) &(km~s$^{-1}$) &(km~s$^{-1}$) && 
}
%\decimalcolnumbers
\startdata
J0241$+$0946 & $-72.32$   & 0.04  &  $ -57.25$ &  0.05 & 45 & $-2.1$ \\
J1144$+$4032 & $-15.22$   & 0.04  &  $  13.81$ &  0.23 & 26 & $-2.2$\\
J1216$+$0231 & $52.473$	  & 1.25  &  $ 106.41$ & 0.69 & 7 &$-2.9:$\\
J1220$+$1637 & $-83.192$  & 0.28  &  $ -38.73$ & 0.33 & 7 &$-3.2$\\
J1231$+$1232 & $32.17$    & 0.09  &  $  14.67$ &  0.16 & 38 & $-2.3$\\
J1859$+$4506 & $-292.42 $ & 0.19  &  $-251.61$ &  0.19 & 14 & $-2.9$\\
J2121$-$0138 & $13.55$    & 0.07  &  $  32.72$ &  0.08 & 41 & $-1.8$\\
\enddata
\end{deluxetable*}

\begin{deluxetable*}{cccccccccc}
%\tablenum{1}
\tablecaption{Equivalent widths of Fe I lines in the primary stars of double-lined spectroscopic binaries \label{tab:ewsb2}}
\tablewidth{0pt}
\tablehead{
\colhead{Wavelength} & \colhead{$\log gf$} & \colhead{L.E.P.} & \colhead{J0241$+$0946} & \colhead{1144$+$4032} &\colhead{J1216$+$0231} &\colhead{J1220$+$1637} &\colhead{J1231$+$1232} & \colhead{J1859$+$4506} & \colhead{J2121$-$0138} \\
({\AA}) & & (eV) & (m{\AA}) & (m{\AA}) & (m{\AA}) & (m{\AA}) & (m{\AA}) & (m{\AA}) & (m{\AA}) 
}
%\decimalcolnumbers
\startdata
4063.594 & 0.062 & 1.558 & 157.6 & \nodata & 81.2 & \nodata & 107.8 & 111.6 & 170.6\\
4071.738 & $-$0.008 & 1.608 & 119.6 & \nodata & 83.8 & 71.2 & 110.5 & 80.1 & 139.9\\
4132.058 & $-$0.675 & 1.608 & 99.8 & 110.0 & 40.8 & \nodata & 71.6 & \nodata & 106.1\\
4181.755 & $-$0.370 & 2.832 & 54.4 & \nodata & \nodata & \nodata & \nodata & 21.6 & 60.9\\
4187.795 & $-$0.554 & 2.425 & 63.6 & \nodata & \nodata & \nodata & \nodata & \nodata & 67.5 \\
4191.430 & $-$0.670 & 2.469 & 46.0 & \nodata & \nodata & \nodata & 65.7 & \nodata & 64.8 \\ 
\enddata
\end{deluxetable*}

\clearpage 


\bibliography{paper1}{}
\bibliographystyle{aasjournal}

%\begin{thebibliography}{}
%\bibitem[Aoki et al.(2015)]{aoki15} Aoki, W., Suda, T., Beers, T.~C., et al.\ 20
%15, \aj, 149, 39
%
%\bibitem[Green et al.(2018)]{green18} Green, G.~M., Schlafly, E.~F., Finkbeiner, %D., et al.\ 2018, \mnras, 478, 651
%
%\bibitem[Kim et al.(2002)]{kim02} Kim, Y.-C., Demarque, P., Yi, S.~K., et al.\ 2
%002, \apjs, 143, 499
%
%\bibitem[Munari \& Zwitter(1997)]{munari97} Munari, U. \& Zwitter, T.\ 1997, \aap, %318, 269
%\end{thebibliography}



\end{document}